\documentclass[twocolumn,prl,showpacs,showkey,nofootinbib,amsmath,amssymb]{revtex4}

\usepackage{graphicx}
\usepackage{dcolumn}
\usepackage{bm}
\usepackage[francais]{babel}
\usepackage{psfrag}
\usepackage{hyperref}

\begin{document}

\preprint{APS/123-QED}

\title{A dynamical law for slow crack growth in polycarbonate films}

\author{Pierre-Philippe Cortet}
\author{Lo\"{i}c Vanel}
\email{Loic.Vanel@ens-lyon.fr}
\author{Sergio Ciliberto}
\affiliation{Laboratoire de physique, CNRS UMR 5672,
  Ecole Normale Sup\'erieure de Lyon, Universit\'e de Lyon,
  46 all\'ee d'Italie,
  69364 Lyon Cedex 07, France}

\date{\today}

\begin{abstract}
We study experimentally the slow growth of a single crack in
polycarbonate films submitted to uniaxial and constant imposed
stress. For this visco-plastic material, we uncover a dynamical
law that describes the dependence of the instantaneous crack
velocity with experimental parameters. The law involves a
Dugdale-Barenblatt static description of crack tip plastic zones
associated to an Eyring's law and an empirical dependence with the
crack length that may come from a residual elastic field.
\end{abstract}


\maketitle

Stressed solids commonly break apart once a critical stress
threshold is reached. However, many experiments
\cite{Zhurkov,Bueche,Brenner,Pauchard,Santucci} show that a given
solid submitted to a subcritical stress breaks after a certain
amount of time. Therefore, understanding the mechanisms of
subcritical rupture of solids has become an important goal of
fracture physics in order to improve the resistance of structures to
delayed failure that may have catastrophic consequences. According
to reported experimental works \cite{Zhurkov,Bueche}, the dependence
of the rupture time with applied stress $\sigma$ can be described in
many kinds of materials (polymers, metal alloys, semi-conductors,
rocks...) by an Arrhenius law with an energy barrier decreasing
linearly with $\sigma$. This proposed universality is disturbing
since these materials have micro-structures and rheological
properties very different from one another, and the rupture dynamics
is certainly expected to be dependent on those properties. To lift
this paradox, one must go beyond characterization of global
properties such as rupture time and instead study experimentally the
full time-resolved rupture dynamics, from the stress application to
the final breakdown of the sample. A convenient system to start with
is a two-dimensional solid with a single macroscopic initial crack
submitted to a uniaxial constant load.

In this context, recent experimental studies \cite{Santucci} have
shown that subcritical crack growth in paper sheets can be
successfully described by a thermally activated mechanism inspired
from previous theoretical works in elastic brittle media
\cite{Hsieh, Santucci3}. Experimental study of slow crack growth in
a visco-plastic material under stress is a very active topic
\cite{Haddaoui}. General theoretical frameworks
\cite{Schapery,Kaminskii,Chud_growth} have been proposed to predict
the dependence of the crack growth velocity with experimental
parameters using characteristic material time-response functions
such as its compliance. However, these models involve complex
integro-differential equations which are hardly tractable in
practical situations where visco-plastic effects are strong.
Consequently, the experimental time evolution of the
\emph{instantaneous} crack growth dynamics can not be captured
easily by current models.

In order to provide more experimental insight in our understanding
of visco-plastic effects during slow crack growth, we have performed
an experimental study of the slow growth of a single crack in
amorphous polymer films made of polycarbonate which is a highly
non-brittle visco-plastic material. The experiments consist in the
growth of a single linear crack in a polycarbonate film submitted to
uniaxial and constant imposed force. The polycarbonate films used
are Bayer Makrofol\textsuperscript{\textregistered} DE and have the
properties of bulk material. Before each experiment, a crack of
length $\ell_i$ (from $0.5$ to $3$cm) is initiated at the center of
the polycarbonate sample (height $21$cm, length $24$cm). Then, a
constant force $F$ is applied to the film perpendicularly to the
crack direction, so that we get a mode 1 crack opening
configuration. Using a camera, we follow the growth of the crack
length $\ell$ under constant applied stress $\sigma=F/eH$ ($e$ is
the film thickness and $H$ the sample height) until the total
rupture of the sample. The applied stress $\sigma$ is chosen such
that crack growth is slow, i.e. smaller than a critical one
$\sigma_c$, above which crack propagation occurs in a few seconds.
More details about the experimental set-up can be found in
\cite{Cortet}.

\begin{figure}
\psfrag{L}[c]{$\ell$} \psfrag{H}[c]{$\ell_{\rm{\tiny pz}}$}
    \centerline{\includegraphics[width=8.5cm]{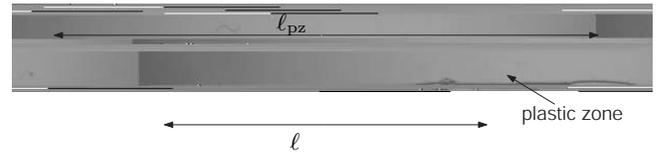}}
    \caption{Image of a crack in a polycarbonate film with its
    macroscopic plastic zone at each tip.} \label{imfrac}
\end{figure}

In each crack growth experiment, during the loading phase of the
film, a macroscopic flame-shaped plastic zone appears at each tip of
the crack \cite{Donald} and grows with the applied stress (cf. Fig.
\ref{imfrac} where is defined the plastic zone length from tip to
tip $\ell_{\rm{\tiny pz}}$). In the late loading stage, the crack
may also start to grow at a time that appears to be statistical. It
is probably a consequence of the dispersion in the local toughness
of the material or in the initial crack tip shape. Consequently, the
real experimental initial condition, obtained when the constant
stress $\sigma$ is reached, is not exactly $\ell=\ell_i$. Depending
on the moment when the crack starts to grow during the loading
phase, the true initial condition of the creep experiment will be a
couple of value for the crack and plastic zone length (Fig.
\ref{imfrac}): ($\ell^{*}$, $\ell_{\rm{\tiny pz}}^{*}$). Finally,
during the imposed stress stage, the plastic zones and the crack are
both growing until the final breakdown of the sample in a way that
the crack never catches up the plastic zone tip. Inside the plastic
zone, the film is subjected to a thinning which brings its thickness
from $125\mu$m to about $75\pm5\mu$m.

\begin{figure}[h]
    \psfrag{Z}[c]{$\ell-\ell_{x}$ (cm)}
    \psfrag{X}[c]{$t$ (s)}
    \psfrag{W}[c]{$t-t_x $ (s)}
    \psfrag{Y}{$t/\tau$}
    \psfrag{L}[c]{$\ell$}
    \psfrag{U}[c]{length (cm)}
    \psfrag{V}[c]{$T_r$}
    \centerline{\includegraphics[width=5.8cm]{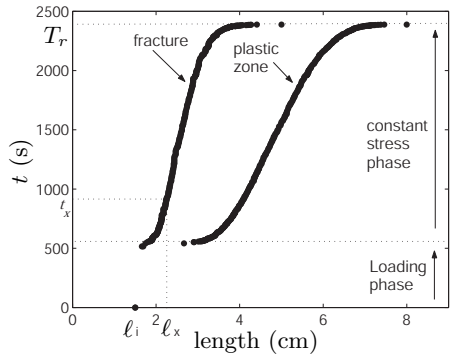}}
    \caption{\label{t_long} Time as a function of both the crack and process
    zone lengths for an imposed stress experiment ($\ell_i=1.5$cm,
    $F=900$N). We indicate the position of the inflexion point $t_x$, $\ell_x$ of the crack growth curve.}
\end{figure}

Typical growth curves of the fracture and plastic zone are shown in
Fig. \ref{t_long}. Both curves show a quite similar smooth shape.
This regular shape lets us think that the crack growth in
polycarbonate films is a deterministic phenomenon. However, for
identical experimental conditions, we notice a large dispersion of
the rupture times and more generally of the crack growth dynamics.
There is actually up to a factor five between the rupture time of
the fastest and slowest experiments. We suggest that the explanation
for this statistics in the crack growth dynamics does not come from
the growth mechanism itself, but is a consequence of the dispersion
in the effective initial conditions at the beginning of the constant
stress phase of the experiment ($\ell^*$, $\ell_{\rm{\tiny pz}}^*$).
These initial conditions are clearly statistical and hardly
controllable in our experiment. They are dependent on the moment
when the crack starts growing during the loading stage of the sample
and they determine all the rest of the experiment.

\begin{figure}[h]
\psfrag{X}[c]{$\sigma$ (N.m$^{-2}$)}
\psfrag{F}[l][][0.8]{$y=-1.56\,10^{-6}x+62.7$}
\psfrag{Z}[c]{$\log\langle T_r \rangle$}
       \centerline{\includegraphics[width=6cm]{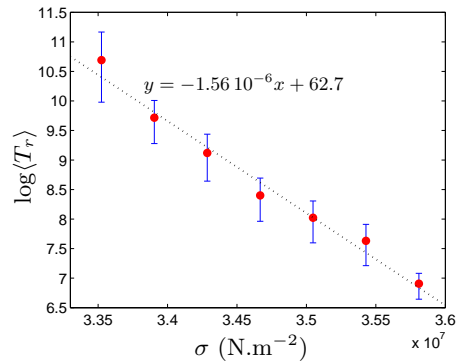}}
       \caption{Natural logarithm of the average rupture time
        as a function of the applied stress for a series of
        experiments performed for $\ell_i=1.5$cm.}
       \label{trupt_1.5}
\end{figure}

In Fig. \ref{trupt_1.5}, we show the evolution of the average
rupture time $\langle T_r \rangle$ (averaged over at least ten
experiments) as a function of the applied stress for a series of
experiments performed at $\ell_i=1.5$cm. We see a linear dependence
of $\log \langle T_r \rangle$ with the applied stress that
corresponds well to an exponential description of the rupture time
as proposed by Zhurkov \cite{Zhurkov}. The linear fit of the data is
of quite good quality and suggests that $\langle T_r \rangle
=T_0\,e^{-a \sigma}$. In Zhurkov's approach, the stress dependence
of $\langle T_r \rangle$ is interpreted as an Eyring's law
\cite{Eyring} with $a=V/k_BT$ where $V$ is assumed to be a
characteristic volume of the material. However, in our experiments,
the parameter $V$ can not be a constant since a different initial
crack length $\ell_i$ gives a completely different rupture time for
the same applied stress. Thus, the external applied stress $\sigma$
can not be the single control parameter of the rupture dynamics.
Then, it is clear that Zhurkov's description needs to be improved to
take into account the specific geometry of the problem. In
particular, the stress $\sigma_y$ holding in the plastic zone close
to the crack tips most probably participates in the dynamical
processes leading to the crack growth.

\begin{figure}
    \psfrag{X}[c]{$\overline{\sigma_{y}}$ (N.m$^{-2}$)}
    \psfrag{Y}[c]{$\log \overline{v}$}
    \psfrag{H}[l][][0.8]{$y=6.27\,10^{-7}x-46.3$}
    \centerline{\includegraphics[width=6cm]{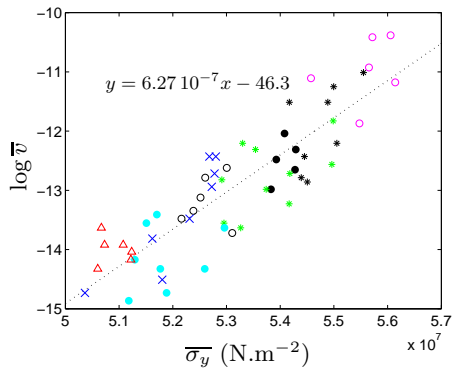}}
    \caption{Natural logarithm of the average crack growth velocity $\overline{v}$
    as a function of the average plastic stress during the growth.
    Each point represents the average dynamical behavior during an
    experiment. Experimental conditions are various ($\ell_i=1.5,2,3$cm and $2.9<\sigma<3.8\,10^{7}$N.m$^{-2}$).
    Each experimental condition corresponds to different symbols.}
    \label{lbdsurtau2}
\end{figure}

The Dugdale-Barenblatt cohesive zone model \cite{Dugdale,Barenblatt}
is a good and simple mean to estimate the stress $\sigma_y$. This
quantity appears intuitively as a possible control parameter for the
crack dynamics just like the stress intensity factor is for brittle
materials. Dugdale-Barenblatt model predicts:
\begin{equation}
\label{dbinvert}
\sigma_y=\frac{\pi}{2}\frac{\sigma}{\rm{arcos}\left(\frac{\ell}{\ell_{\rm{\tiny
pz}}}\right)}.
\end{equation}
This plastic stress $\sigma_y$ can be computed at each moment using
Eq.~(\ref{dbinvert}) with the instantaneous values of $\sigma$,
$\ell$ and $\ell_{\rm{\tiny pz}}$. To account for the global
dynamics during an experiment, we compute the time-averaged growth
velocity on the whole experiment $\overline{v}$ and compare it to
the time-averaged plastic stress $\overline{\sigma_y}$ (see Fig.
\ref{lbdsurtau2}). Each point of this Figure represents the mean
behaviour over an experiment. The data are compatible with a linear
law that predicts an exponential dependence of the average growth
velocity with the mean stress in the plastic zone:
\begin{equation}\label{vmean_sigy}
\overline{v}=v_0\,e^{a \overline{\sigma_y}}
\end{equation} with
$a=6.3\,10^{-7}$m$^2$.N$^{-1}$ and $v_0=7.8\,10^{-21}$m.s$^{-1}$.

It is striking that the prefactor of the stress in the exponential
curve (cf. Eq.~(\ref{vmean_sigy})) is close quantitatively to the
one obtained in the Eyring's law for the polycarbonate creep
\cite{Cortet}. Both prefactors probably correspond to a unique
material constant $V/k_BT$. Thus, we can conclude that the Eyring's
law plays a central role in the mechanisms of crack growth in
polycarbonate films.

\begin{figure}
\psfrag{A}{(a)} \psfrag{B}{(b)}
    \psfrag{X}[c][][0.9]{$\sigma_{y}$ (N.m$^{-2}$)}
    \psfrag{t}[c][][0.9]{time}
    \psfrag{Y}[c][][0.9]{$\log v$}
    \psfrag{Z}[c][][0.9]{$\sigma_{y}^{\rm{\tiny corr}}$ (N.m$^{-2}$)}
    \psfrag{G}[r][][0.8]{$y=6.83\,10^{-7}x-48.8$}
    \centerline{\includegraphics[width=6cm]{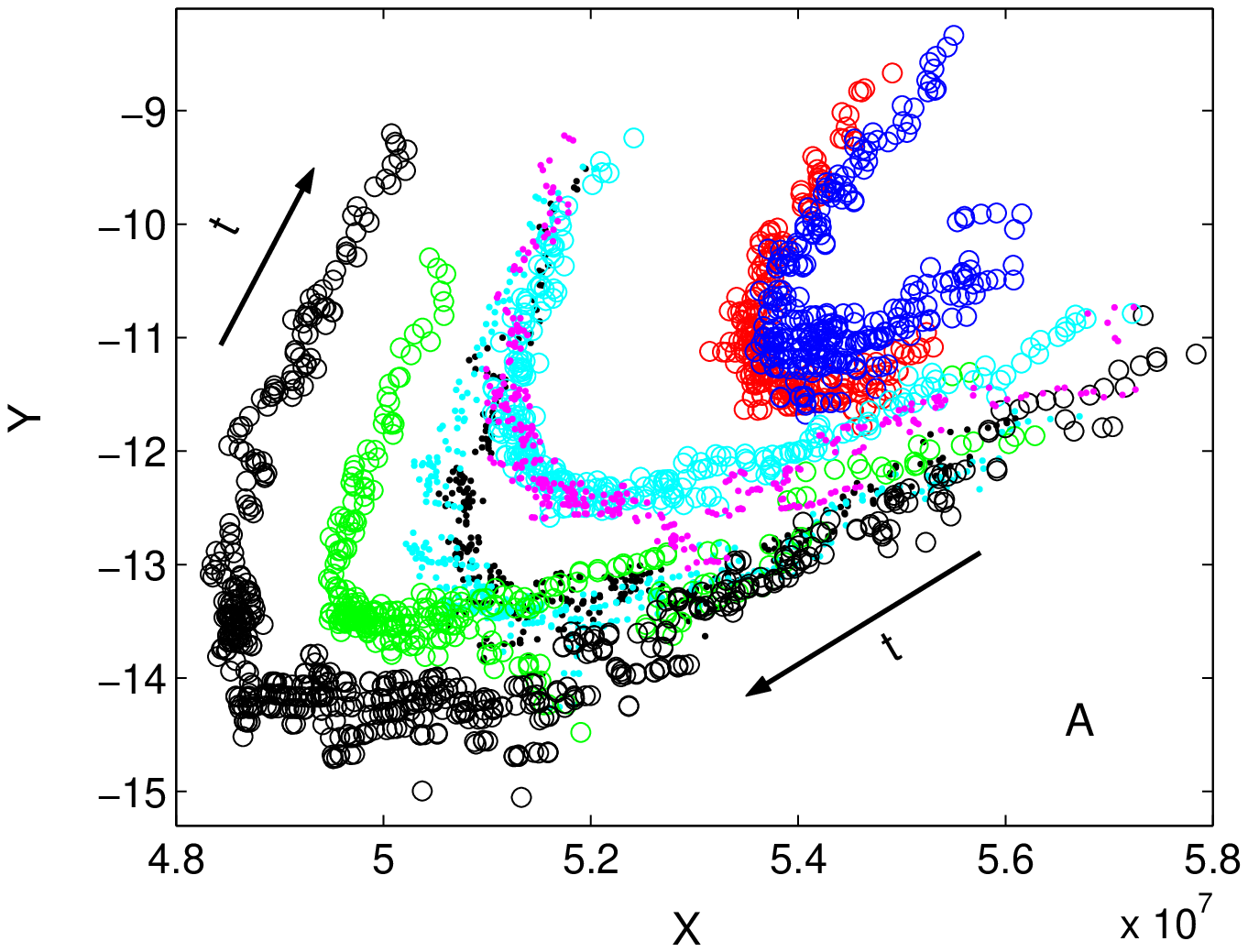}}\vspace{0.2cm}
    \centerline{\includegraphics[width=5.7cm]{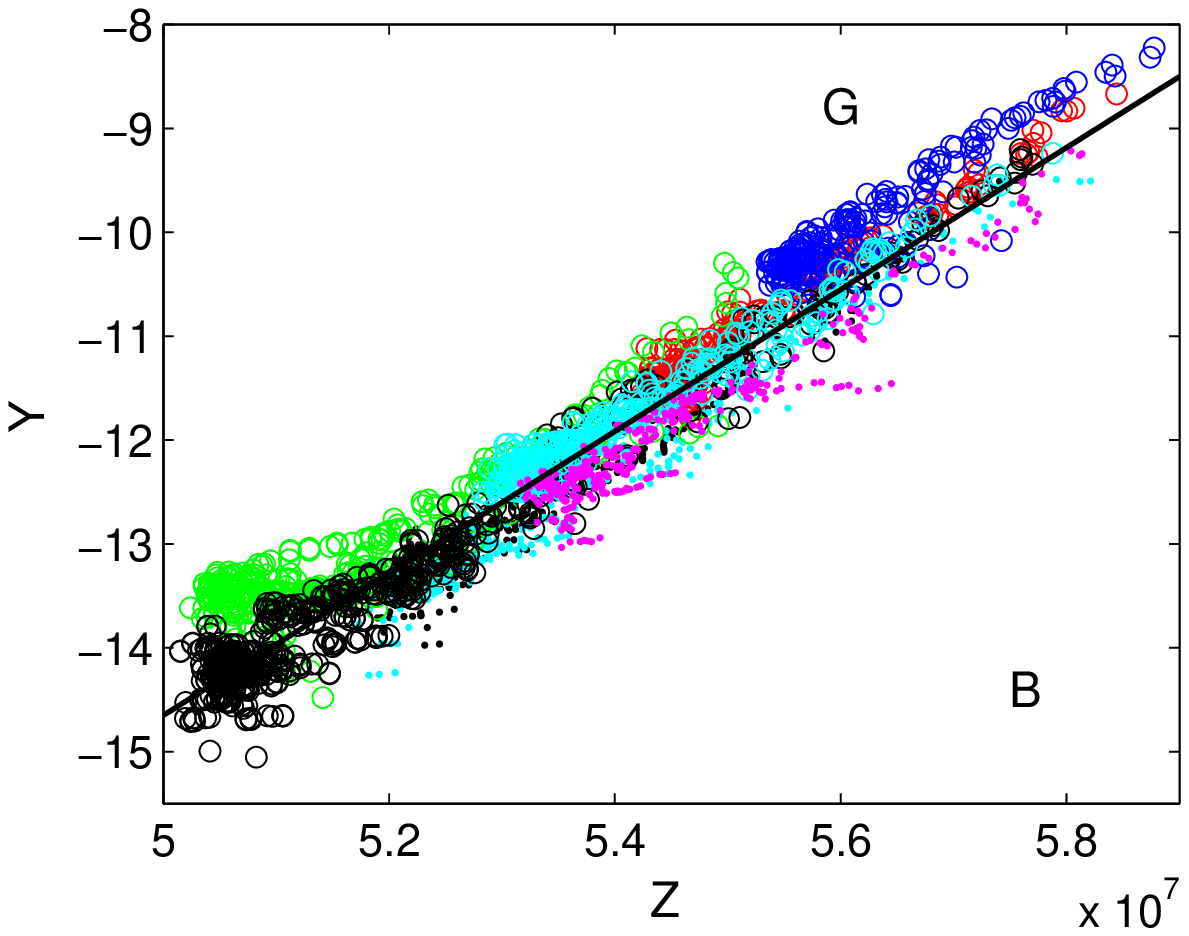}}
    \caption{Natural logarithm of the instantaneous crack growth velocity as a function of
    (a) the Dugdale-Barenblatt stress, (b) the corrected Dugdale-Barenblatt stress
    $\sigma_y^{\rm{\tiny corr}}$ according to Eq. (\ref{DBcorr}) for height experiments
    performed with various experimental conditions
    ($\ell_i=1.5,2,3$cm and $2.9<\sigma<3.8\,10^{7}$N.m$^{-2}$). In Fig. (b),
    the black line is the result of a linear data fit.}
    \label{vit_sigy}
\end{figure}

We now go beyond a simple analysis of the average growth dynamics
by looking at the dependence of the crack velocity with the stress
in the plastic zone at each time during the crack growth. We plot
in Fig. \ref{vit_sigy}(a) the instantaneous crack velocity
$v=d\ell/dt$ as a function of the instantaneous value of the
Dugdale-Barenblatt stress $\sigma_y$ for height experiments
performed with various experimental conditions. Here, the
description of the instantaneous velocity by an exponential law
fails, especially when the crack length becomes larger than
$\ell_x$ at which the minimum crack velocity is reached ($\ell_x$
is also the inflexion point of the growth curve in Fig.
\ref{t_long}). In fact, the Eyring's law given by Eq.
(\ref{vmean_sigy}) describes well the behaviour only when $\ell
\simeq \ell_x$. We discovered that introducing a correction to
$\sigma_y$ linear with the crack length $\ell$ allows us to
collapse the experimental data on a straight line (cf. Fig.
\ref{vit_sigy}(b)). This correction can be written as:
\begin{equation}\label{DBcorr}
\sigma_y^{\rm{\tiny
corr}}=\frac{\pi}{2}\frac{\sigma}{\rm{arcos}\left(\frac{\ell}{\ell_{\rm{\tiny
pz}}}\right)}+\kappa\,(\ell-\ell_x).
\end{equation} For each experiment, we
determine the value $\kappa= (3.4 \pm 0.6) \,10^8$N.m$^{-3}$. The
dispersion of $\kappa$ values seems to be statistical as no
systematic dependence with $\sigma$ or $\ell_i$ could be found.
This rescaling means that the crack growth velocity seems to
follow:
\begin{equation}\label{crpcbbis}
\frac{d\ell}{dt}=v_0\,e^{\frac{V}{k_BT}\sigma_y^{\rm{\tiny corr}}}.
\end{equation}
The collapse of the data for various experimental conditions means
that $v_0$ can be considered as a constant.

\begin{figure}
    \psfrag{U}[l][][0.8]{$y=-1.01\,x+4.2\,10^7$}
    \psfrag{V}[l][][0.8]{$y=-3.57\,10^8\,x+4.07\,10^7$}
    \psfrag{Y}[c][][0.9]{$\kappa \ell_x$ (N.m$^{-2}$)}
    \psfrag{W}[c][][0.9]{$\sigma$ (N.m$^{-2}$)}
    \psfrag{X}[c][][0.9]{$\sigma_c$ (N.m$^{-2}$)}
    \psfrag{Z}[c][][0.9]{$\ell_i$ (m)}
    \psfrag{A}{(a)}
    \psfrag{B}{(b)}
    \centerline{\includegraphics[width=5.4cm]{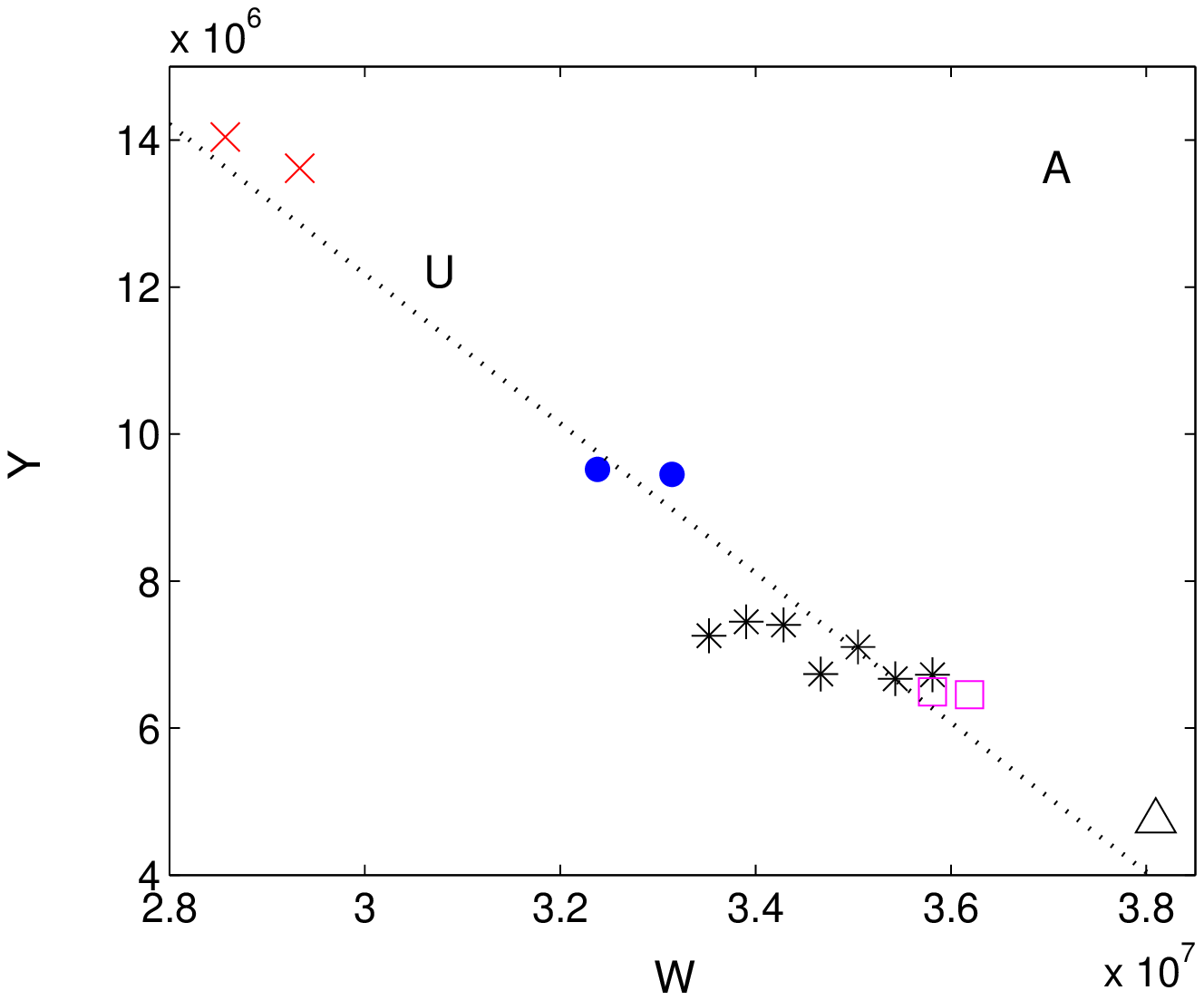}}\vspace{0.2cm}
    \centerline{\includegraphics[width=6cm]{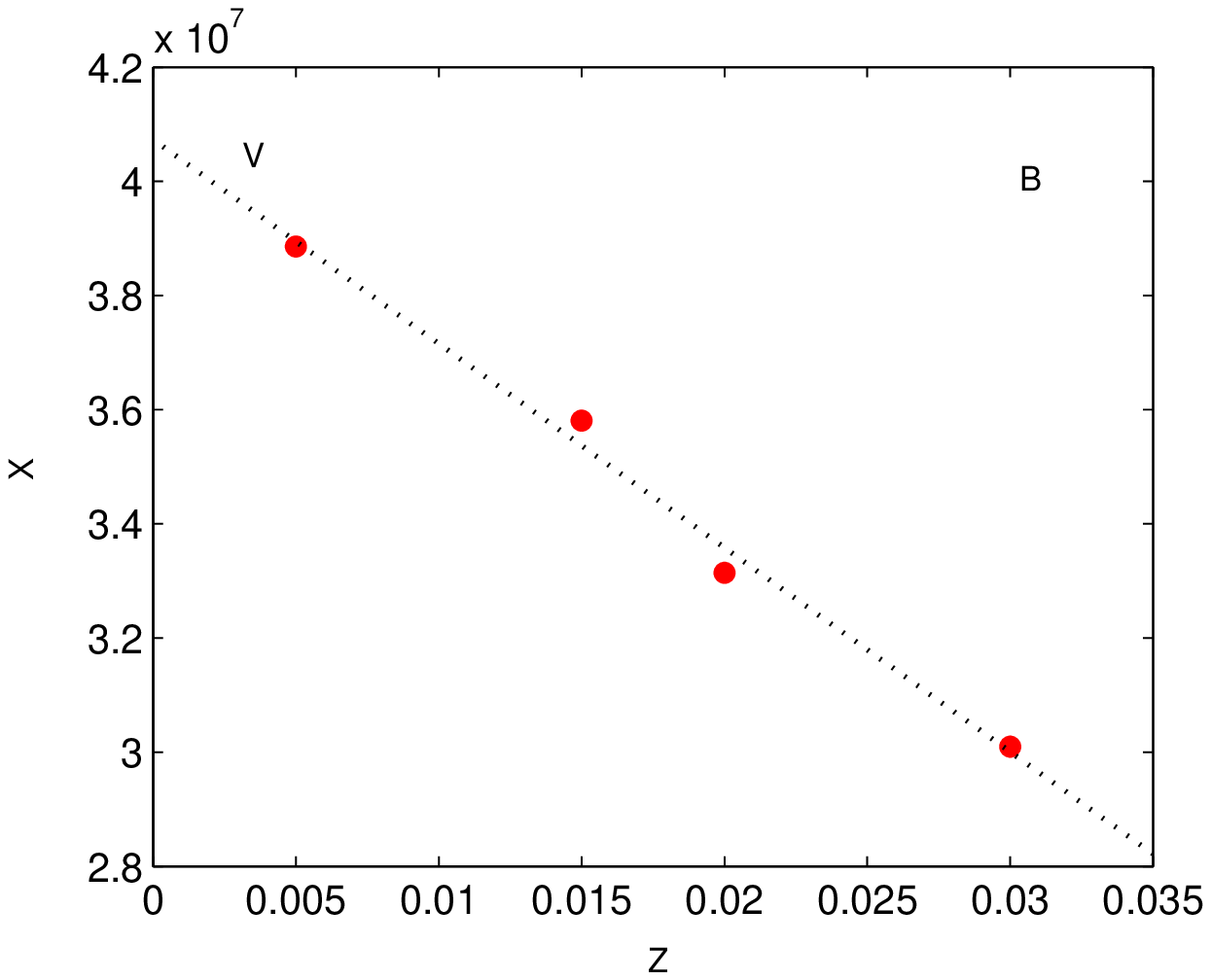}}
    \caption{\small (a) $\kappa \ell_x$  for various experimental conditions
    ($\ell_i=1.5,2,3$cm and $2.9<\sigma<3.8\,10^{7}$N.m$^{-2}$) as a function of the applied stress
    $\sigma$. (b) Critical rupture stress $\sigma_c$ as a function of the initial crack length
    $\ell_i$. The dotted lines are a linear fit of the data.}
    \label{lx_sigma}
\end{figure}

In Eq. (\ref{DBcorr}), the crack length at the inflexion point in
the growth curve plays a particular role. It turns out that its
value depends on the experimental conditions. This can be seen in
Fig.~\ref{lx_sigma}(a) where the product $\kappa \ell_x$, with
$\kappa=3.4 \,10^8$N.m$^{-3}$, is plotted as a function of the
applied stress $\sigma$. Remarkably, the dependence of $\kappa
\ell_x$ with $\sigma$ is well approximated by a linear relation:
$\kappa \ell_x = \sigma_{x}-\sigma$, where $\sigma_{x}=4.2\,10^{7}
$N.m$^{-2}$.

A way to clarify the meaning of this relation is to look at the
dependence of the critical stress $\sigma_c$ needed to break
instantaneously a sample with a crack of initial length $\ell_i$. In
brittle materials, we would expect this critical stress to decrease
in $1/\sqrt{\ell_i}$ since the rupture criterion is reached when the
initial stress intensity factor equals the toughness of the material
$K_c$ \cite{Santucci}: $\sigma_c \sqrt{\pi\ell_i/2}=K_c$. For an
amorphous visco-plastic material such as polycarbonate, we do not
get the same functional dependence. Indeed, as we can see in
Fig.~\ref{lx_sigma}(b), the relation between $\sigma_c$ and $\ell_i$
can be approximated by a linear equation: $\beta
\ell_i=\sigma_s-\sigma_c$, where $\sigma_{s}=4.07\,10^{7}
$N.m$^{-2}$ and $\beta=3.57\,10^8$N.m$^{-3}$. We note that $\kappa
\simeq \beta$ and $\sigma_x \simeq \sigma_s$ and will consider these
quantities to be the same material constants. So, we find that the
quantity $\Sigma(\sigma_c,\ell_i)=\sigma_c+\kappa \ell_i$ may play a
similar role than the initial stress intensity factor in brittle
materials. Furthermore, it allows us to interpret the value of the
crack length at the inflexion point as defined by a characteristic
value of the quantity $\Sigma(\sigma,\ell_x)=\sigma_{x}\simeq
\sigma_{s}$ that corresponds to an intrinsic property of
polycarbonate. Indeed, $\sigma_{s}$ corresponds to the rupture
threshold $\sigma_c$ in the limit when there is no initial crack.

According to the previous analysis of the instantaneous crack
velocity, crack growth in polycarbonate films appears to be ruled,
during an experiment, by an Eyring's law (cf. Eq (\ref{crpcbbis}))
with:
\begin{equation}\label{DBcorr2}
\sigma_y^{\rm{\tiny
corr}}=\frac{\pi}{2}\frac{\sigma}{\rm{arcos}\left(\frac{\ell}{\ell_{\rm{\tiny
pz}}}\right)}+\kappa\,\ell+\sigma-\sigma_s
\end{equation}
This effective stress $\sigma_y^{\rm{\tiny corr}}$ is composed of
the Dugdale-Barenblatt estimation of the crack tip plastic zone
stress $\sigma_y$, a linear dependence with the crack length
$\kappa\, \ell$ and the applied stress at the borders of the sample
$\sigma$. Note that in Eq. (\ref{crpcbbis}) appears naturally a
volume $V \simeq 2.8\,10^{-27}$m$^{3}$ close to the one used to
describe the simple creep flow of polycarbonate ($3.1
\,10^{-27}$N.m$^{-2}$) as well as the growth of a necking
instability in polycarbonate films ($3.0 \,10^{-27}$m$^{3}$). This
observation reinforces the idea that the Eyring's law for crack
growth is truly a consequence of the creep behavior of
polycarbonate.

In Eq. (\ref{DBcorr2}), the viscous relaxation is taken into account
by the experimentally measured evolution of the ratio
$\ell/\ell_{pz}$ as the crack grows. Indeed, if this ratio was
constant, the stress in the plastic zone would also be constant and
the velocity would increase monotonously due to the linear term in
crack length. In that case, the behavior would actually be
qualitatively the same as the one for crack growth in brittle
facture \cite{Santucci}. To predict fully the \emph{viscous}
dynamics of the crack, we need a second equation that will prescribe
$\ell_{\rm{\tiny pz}}$:
\begin{equation}\label{vitlpz}
\frac{d\ell_{\rm{\tiny pz}}}{dt}=f(\ell_{\rm{\tiny
pz}},\ell,\dot{\ell},\sigma,...)
\end{equation}
An original theoretical approach recently developed by Bouchbinder
\cite{Bouchbinder} in extension to the Shear-Transformation-Zone
Theory proposed by Falk and Langer \cite{Falk} is certainly useful
for deriving an equation of the plastic zone velocity (cf. Eq.
(\ref{vitlpz})). Additionally, numerical simulations that can
reproduce the complex visco-plastic behavior of polycarbonate may
help in going further in the interpretation of our experimental
results \cite{Boyce,Gearing,Estevez}.

\end{document}